\documentclass[aps,twocolumn]{revtex4}
\usepackage{amssymb}
\usepackage{amsfonts}
\usepackage{amsmath}
\usepackage{graphicx}
\begin{document}
\preprint{ }
\title{Current reversals in rapidly rotating ultra-cold Fermi gases}
\author{K. Bencheikh and  S. Medjedel}
\affiliation{Laboratoire de Physique Quantique et Syst\`{e}mes Dynamiques, D\'{e}partement de physique, Universit\'{e} Ferhat Abbas S\'{e}tif,  S\'{e}tif 19000, Algeria.}
\author{G. Vignale}
\affiliation{Department of Physics and Astronomy, University of Missouri, 
Columbia, Missouri 65211, USA.}
\date{\today}

\begin{abstract}
We study the equilibrium current density profiles of harmonically trapped ultra-cold Fermi gases in quantum Hall-like states that appear when the quasi-two-dimensional trap is set in fast rotation.  The density profile of the gas (in the rotating reference frame) consists of incompressible strips
of constant quantized density separated by compressible regions in which the density varies.  Remarkably, we find that the atomic currents flow in opposite directions
in the compressible and incompressible regions -- a prediction that should be amenable to experimental verification.   
\end{abstract}
\maketitle
\begin{figure}
\begin{center}
\includegraphics[width=1.0\linewidth]{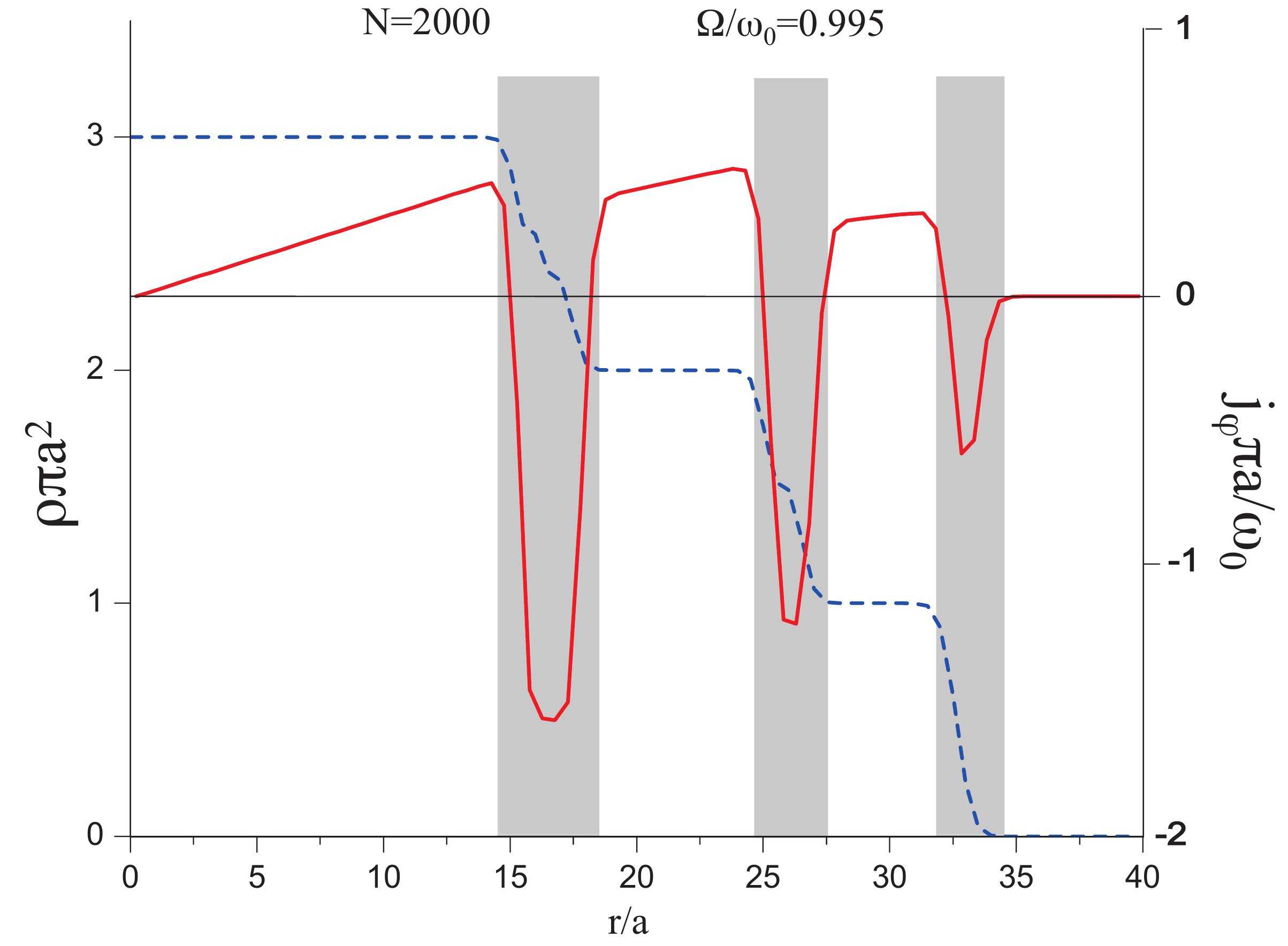}
\end{center}
\caption{(Color online) Particle and current density profile for $N=2000$ atoms in a parabolic trap with characteristic frequency $\omega_0$  rotating at frequency $\Omega$, with $\Omega/\omega_0=0.995$. The radial coordinate $r$ from the center of the trap is in units of $a=\sqrt{\frac{\hbar}{M^*\omega_0}}$, which practically coincides with $\ell$ at this value of $\Omega$.  The dashed blue line represents the particle density, $\rho$, whose value is given on the left scale in units of $\frac{1}{\pi a^2}$.  The solid red line represents the azimuthal component of the particle current density $j_\phi$, whose value is given on the right scale in units of $\frac{\omega_0}{\pi a^2}$.  In the incompressible regions (shown on white background) the density is quantized to an integer, and the current is positive.  In the compressible regions (shown on grey background) the density is un-quantized and the current is negative.}
\label{Fig1}
\end{figure}

Starting with the discovery of BEC in trapped ultra-cold gases of bosonic atoms, the field of  quantum gases has experienced an explosive growth~[1-11].  Indeed, cold atomic systems offer the opportunity to study, in clean and controlled conditions,  key concepts of condensed matter theory, and also to simulate, via the application of suitable optical fields, interactions (such as spin-orbit interactions, periodic potentials) which,  in a conventional solid state environment, arise from microscopic, hardly controllable processes.  One of the fields that can be artificially simulated in cold atoms is the one arising by the fast rotation of the ``vessel" that contains the particles [11].  It is well known~\cite{LandauLifshitzV} that, in the rotating frame, the hamiltonian of the system is modified by the addition of the field $H_{rot}=-\vec \Omega\cdot {\vec L}$, where $\vec \Omega$ is the angular velocity of the ``vessel", and ${\vec L}$ is the operator of the total angular momentum of the system.  This is similar to the coupling of a charged particle to a magnetic field $\vec B$, but the diamagnetic term (quadratic in $\vec B$) is absent.  The response of bosonic or fermionic atomic gases to such a ``rotation field" has been a subject of intensive investigations during the last decade.  As the frequency of rotation increases, an increasing number of vortices appears in a Bose Einstein condensate (BEC).  In the fast-rotation limit, defined by the condition $\Omega \gg \frac{\hbar \rho}{2M^*}$, where $\rho$ is the (two-dimensional) atomic density and $M^*$ is the mass of an atom,  the system exhibits atomic quantum Hall states and the BEC can be described in terms of Landau-level single-particle states [1-7].  More recently,  theoretical studies on rotating Fermi gases have confirmed the existence of Landau-like energy levels [8] and have shown how the Fermi statistics imprints the structure of these energy levels in the density profiles [7-10].   
In the limit of fast-rotating trap these density profiles are found to consist of a series of density plateaus  at quantized densities $\rho= \frac{\nu}{\pi \ell^2}$, where $\nu$ is an integer and $\ell^2 = \frac{\hbar}{M^* \Omega}$ is the square of the ``magnetic length", separated by compressible regions, in which the density varies (see Fig.~\ref{Fig1}).   These findings are in perfect agreement with corresponding results for the density of a two-dimensional electron gas at high magnetic field\cite{[12],Chklovski92,Geller-Vignale}.   Each plateau corresponds to the occupation of a new pseudo-Landau-level  (see Fig.~\ref{Fig2} below).   In addition,  the angular momentum of the gas has been shown to display quantum oscillations~\cite{Grenier2013} as a function of particle number, which are analogous to the de Haas van Alphen oscillations~\cite{Alphen32} of the magnetization in solid state systems.

The importance of the compressible regions emerges when one considers the mass current density profile, $\vec j(\vec r)$ in the rotating frame.  This is a measurable property that we calculate here without approximations.    The existence of equilibrium currents -- also known as persistent currents --  is a purely quantum-mechanical phenomenon, which has been widely studied in  mesoscopic electronic system subjected to magnetic fields~\cite{Buttiker83,Levy90,Chandrasekhar91,Mailly93} or set in rapid rotation~\cite{Vignale-Mashhoon}. Exploiting the analogy between the motion of charged particles in a magnetic field and neutral atoms in a rotating frame, one expects persistent mass current to arise in response to the rotation of an ultra-cold atomic gas.   Our calculations show that the persistent current in the compressible regions flows in a direction {\it opposite to} that of the persistent currents in the incompressible regions.  

There is a good reason for this, since the  nature of the persistent current is quite different in the two types of  regions.  In the incompressible regions it is a Hall-like current, proportional to the gradient of the effective potential acting on the particles, which must be positive to ensure confinement.    By contrast, in the compressible regime we find an edge current, proportional to the gradient of the particle density, which is negative since the density decreases from the center of the trap towards the periphery.   The reversal of the sign of the current in going from an incompressible region to the neighboring compressible one can be viewed as a manifestation of the oscillatory behavior of the local pseudo-magnetization $\vec M(\vec r)$, which is related to the current by the standard relation 
\begin{equation}\label{DefM}
\vec j (\vec r)= \vec \nabla \times \vec M (\vec r)\,.
\end{equation}
The possibility of this representation follows immediately from the
continuity equation $\vec \nabla\cdot\vec j(\vec r)=0$, and the magnetization $\vec M(\vec r)$ is defined up to an arbitrary curl-free field.   The oscillations of $\vec M(\vec r)$ are completely analogous to the  deHaas-van Alphen oscillations of the magnetization of an electronic system, with the difference that our system is charge-neutral, and there is no magnetization: yet the pseudo-magnetization behaves exactly as the true magnetization of a system of charged particles!  The existence of a local pseudo-magnetization satisfying Eq.~(\ref{DefM}) also implies a quantization of the total (integrated) persistent current - which is reminiscent of  the quantization of the vortex current in the superfluid bosonic system.
We believe that our predictions on the current density profile are accessible to experimental verification more easily in this system than in conventional semiconductor-based electron liquids  at high magnetic field.  Spatially resolved Doppler velocimetry techniques can be employed to detect the direction of flow of the atoms.  A detailed verification of this intriguing prediction will thus be possible for the first time.  


Consider a system of $N$ non-interacting Fermions of mass $M^*$ moving in a two-dimensional parabolic trap potential $V(r)=\frac{1}{2}M^* \omega_0^2r^2$, where $r^2=x^2+y^2$.  When the trap is set in rotation with angular frequency $\vec \Omega$ the Hamiltonian in the rotating frame takes the form
\begin{equation}\label{Hamiltonian}
H=\frac{1}{2M^{\ast }}\left(\vec{p}-M^{\ast }\vec{\Omega }\times \vec r\right) ^{2}+V_{eff}(r)\,, 
\end{equation}
with 
\begin{equation}\label{VEFF}
V_{eff}(r)=\frac{M^{\ast }}{2}\left(\omega _{0}^{2}-\Omega
^{2}\right) r^2\,.
\end{equation}
Eq.~(\ref{Hamiltonian}) shows the equivalence of our system to a system of charged particles  in a harmonic trap with characteristic frequency $\omega_{eff}=\sqrt{\omega _{0}^{2}-\Omega^{2}}$ subjected to
uniform magnetic field $B$ such that the cyclotron frequency of the equivalent
magnetic problem is  $\omega _{c}=\frac{eB}{M^*} = 2\Omega$.

\begin{figure}
\begin{center}
\includegraphics[width=1.0\linewidth]{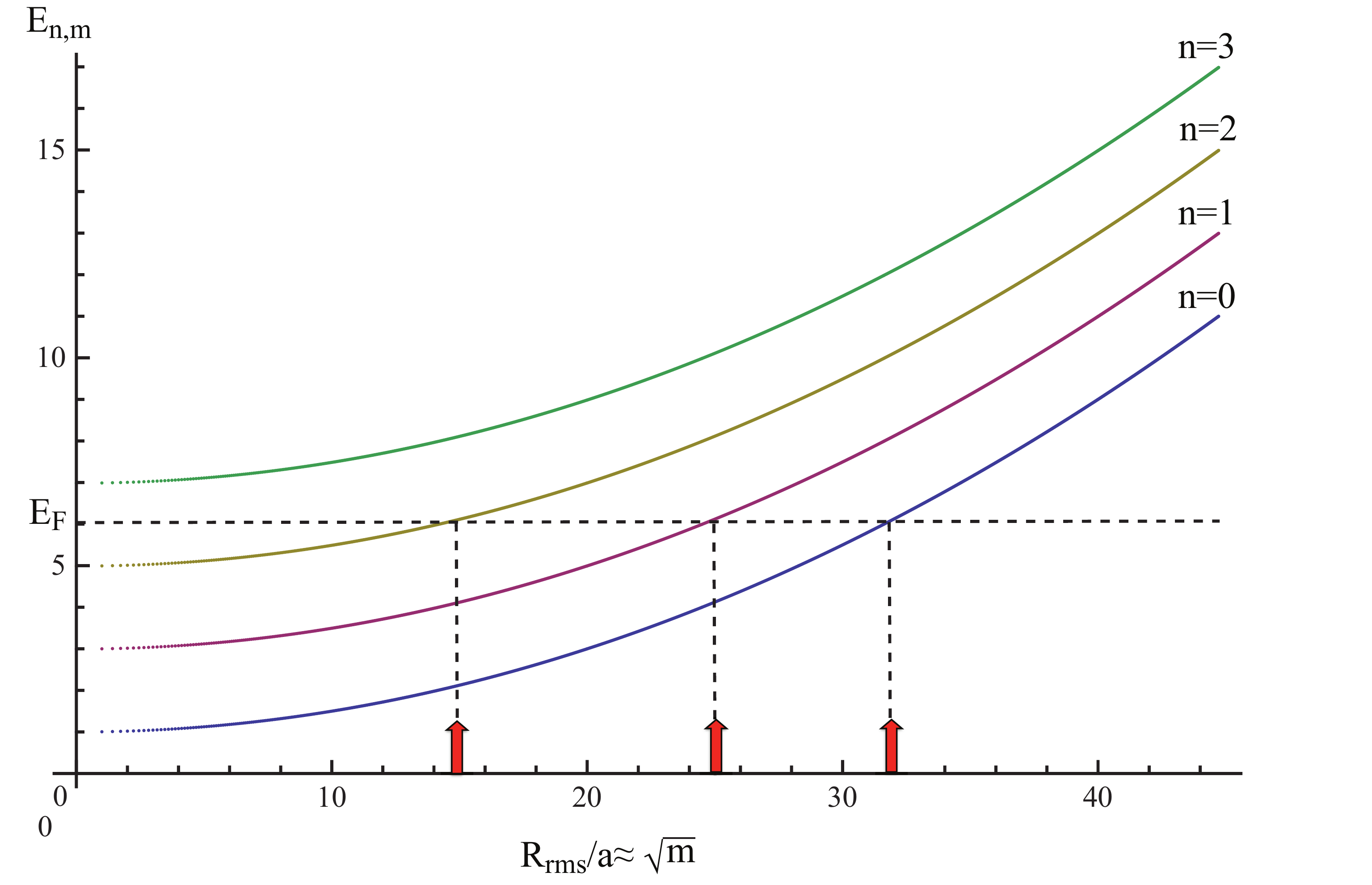}
\end{center}
\caption{(Color online) Energy spectrum in units of $\hbar \omega_0$ for the four lowest lying Landau levels in a rotating two-dimensional trap with $\Omega/\omega_0=0.995$.  Plotted on the horizontal axis is  the root-mean-square distance of the orbital with quantum numbers $n$ and $m$ from the center of the trap: $R_{rms}/a =\sqrt{m}$ for large $m$.   The Fermi energy, $E_F$, is also shown for $N=2000$ atoms.  The positions at which the energy spectrum crosses the Fermi energy are marked by vertical arrows and closely correspond to the centers of the transitions between different incompressible regions shown in Fig.~\ref{Fig1}.}
\label{Fig2}
\end{figure}

The density and the current density are both computed from the single-particle density matrix,
which at the absolute zero of temperature has the form 
\begin{equation}\label{DensityMatrix}
\rho (\vec r,\vec r',\mu)=\sum_{n,m}\phi _{n,m}(\vec r)\phi _{n,m}^{\ast }(\vec r')\Theta \left(\mu-E_{n,m}\right) 
\end{equation}
where $\Theta (x)=1$ or $0$ if $x>0$ or $<0$ respectively, $\mu$ is the Fermi
energy (chemical potential at zero temperature) and  $\phi _{n,m}(\vec r)$ are the eigenfunctions of the Schr\"odinger equation with Hamiltonian~(\ref{Hamiltonian}), and $E _{n,m}$ are the corresponding eigenvalues,
\begin{eqnarray}\label{Eigenvalues}
&&\bigskip E_{n,m}=\hbar (\omega _{0}+\Omega )(n+1/2)+\hbar (\omega_{0}-\Omega )(m+1/2)\nonumber\\
&&n,m=0,1,2....
\end{eqnarray}
In the above energy spectrum the frequency $\omega _{0}$ appears as a
critical value of $\Omega$ since the stability of the gas requires that $
\Omega \leqslant \omega _{0}$ (in the absence of any additional repulsive
potential). In the limit, $\Omega =$ $\omega _{0}$ the centrifugal force,
varying as $M^{\ast }\Omega ^{2}\vec r$\ is balanced by the
trapping force $-M^{\ast }\omega _{0}^{2}\vec r$\textbf{\ }and
the energy spectrum reduces to $E_{n,m}=\hbar \omega _{0}(2n+1)$
which is a Landau-level-like energy spectrum, where the quantum
number $n$ identifies each level and $m$ labels the degenerate states within
each Landau level.  Notice that $m-n$ is the angular momentum of the states in the  Landau
level defined by $n$.

Fig.~\ref{Fig2} shows the energy spectrum $E_{n,m}$ for the four lowest-lying Landau levels ($n=0 - 3$) and $m$ ranging from $0$ to $2000$.   We actually plot the energies versus $\sqrt{m}$, since, for large $m$,  this is the root-mean-square distance of the corresponding orbital from the center of the trap (expressed in units of $a$).  Thus, Fig.~\ref{Fig2} can be viewed as a plot of the orbital energy as a function of its average distance from the center of the trap.  
The particle density 
\begin{equation}\label{Density}
\rho (\vec{r
})=\rho (\vec r,\vec r,\mu )
\end{equation} 
obeys the normalization condition $\int \rho (\vec r)d\vec r
=N$, which determines the Fermi energy $\mu$ as a function of $N$.
In Fig.~\ref{Fig2}  we show the position of the Fermi level for $N=2000$, and $\Omega/\omega_0=0.995$.   Moving from the center to the periphery of the trap we see that the number of occupied Landau levels changes from $3$ to $2$ to $1$, to none.  The transitions occur at the positions for which a branch of the spectrum crosses the Fermi energy.
These positions, marked by vertical arrows in Fig.~\ref{Fig2}, closely correspond to the centers of the transitions between different incompressible regions in Fig.~\ref{Fig1}.  

The particle current density is given by
\begin{equation}\label{CurrentDensity}
\vec{j}(\vec r)=\frac{\hbar }{2M^{\ast }i}\left[
(\vec{\nabla }_{\vec r}-\vec{\nabla }_{
\vec r'})\rho (\vec r,\vec r',\mu )\right] _{\vec r'=\vec r}-\left( 
\vec{\Omega }\times \vec r\right) \rho \left( 
\vec r\right)  
\end{equation}

The density matrix (\ref{DensityMatrix}) can be obtained through an inverse Laplace
transformation,
\begin{equation}\label{LaplaceTransform}
\rho (\vec r,\vec r',\mu)=\frac{1}{2\pi i}
\int\limits_{c-i\infty }^{c+i\infty }d\beta \frac{e^{\beta \mu}
}{\beta }C(\vec r,\vec r';\beta)
\equiv {\cal L}
_{\mu}^{-1}\left[\frac{C(\vec r,\vec r'
;\beta )}{\beta }\right]\,, 
\end{equation}
where $C(\vec r,\vec r';\beta )=\sum_{n}\phi
_{n}(\vec r)\phi _{n}^{\ast }(\vec r')\exp
(-\beta \varepsilon _{n})$ is the so-called Bloch propagator of the
hamiltonian $H$~\cite{[14],[15]}  and where, to carry out the complex integration, $
\beta $ is taken to be a complex variable.
%
%
The Bloch propagator of charged particle in a 2D harmonic trap and subjected
to a magnetic field normal to its surface is exactly known~\cite{[16],[18]}.
\begin{widetext}
Exploiting the equivalence between rotations and magnetic field, we can immediately write down the propagator of  the Hamiltonian~(\ref{Hamiltonian}) :
\begin{eqnarray}
C\left( \vec r,\vec r';\beta \right) &=&\frac{
M^{\ast }\omega _{0}}{2\pi \hbar \sinh \left( \beta \hbar \omega _{0}\right) 
}\text{exp}\left[ -i\frac{M^{\ast }\omega _{0}}{\hbar }\left( xy^{\prime
}-yx^{\prime }\right) \frac{\sinh \left( \beta \hbar \Omega \right) }{\sinh
\left( \beta \hbar \omega _{0}\right) }\right] \times  \notag \\
&&\exp \left[ -\frac{M^{\ast }\omega _{0}}{4\hbar }\left[ (x+x^{\prime
})^{2}+(y+y^{\prime })^{2}\right] \left[ \coth \left( \beta \hbar \omega
_{0}\right) -\frac{\cosh \left( \beta \hbar \Omega \right) }{\sinh \left(
\beta \hbar \omega _{0}\right) }\right] \right] \times  \notag \\
&&\exp \left[ -\frac{M^{\ast }\omega _{0}}{4\hbar }\left[ (x-x^{\prime
})^{2}+(y-y^{\prime })^{2}\right] \left[ \coth \left( \beta \hbar \omega
_{0}\right) +\frac{\cosh \left( \beta \hbar \Omega \right) }{\sinh \left(
\beta \hbar \omega _{0}\right) }\right] \right] 
\end{eqnarray}
The particle density is then given by
an inverse Laplace  transformation, which  is carried out by the method of Ref.~\cite{[17]}.   The final result can be written as
\begin{equation}\label{DensityExplicit}
\rho (\vec r)=\frac{1}{\pi a^{2}}e^{-\frac{r^2}{a^{2}}}\left\{ \sum\limits_{k=0}^{\infty }\sum\limits_{n=0}^{\infty
}\sum\limits_{m=0}^{\infty }\frac{1}{n!m!}\left(r^2/a^{2}\right) ^{n+m}L_{k}^{n+m}\left( 2r^2/a^{2}\right) \Theta \left[ \mu -2k\hbar \omega _{0}-E_{n,m}\right]
\right\}\,,
\end{equation}
where $L_{k}^{\alpha}$ denotes the generalized Laguerre polynomial~\cite{[19],[20]}, and  $a\equiv\sqrt{\frac{\hbar}{M^*\omega_0}}$.  Notice that $a$ coincides with $\ell$ in the limit of fast rotation, $\Omega \to \omega_0$.

Similarly, the particle current density~(\ref{CurrentDensity})  is purely azimuthal $\vec j(\vec r)= j(r)\vec e_\phi$ where $\vec e_\phi$ is the unit vector in the azimuthal direction, and $j(r)$ is given by the formula  
\begin{eqnarray}\label{CurrentDensityImplicit}
j(r) &=&\frac{r}{\pi a^{2}}{\cal L}_{\mu
}^{-1}\left\{ \frac{e^{-\frac{r^2}{a^{2}}\coth \left(
\beta \hbar \omega _{0}\right) }}{\beta }\left[ \frac{\omega _{0}\left( e^{
\frac{\beta \hbar \Omega }{2}}-e^{-\frac{\beta \hbar \Omega }{2}}\right) }{
\left( e^{\frac{\beta \hbar \omega _{0}}{2}}-e^{-\frac{\beta \hbar \omega
_{0}}{2}}\right) ^{2}}-\frac{\Omega }{\left( e^{\frac{\beta \hbar \omega _{0}
}{2}}-e^{-\frac{\beta \hbar \omega _{0}}{2}}\right) }\right] \times \right. 
\notag \\
&&\left. \left[ e^{\left( \frac{e^{\beta \hbar \Omega }}{(e^{\beta \hbar
\omega _{0}}-e^{-\beta \hbar \omega _{0}})}\frac{r^{2}}{
a^{2}}\right) }\times e^{\left( \frac{e^{-\beta \hbar \Omega }}{(e^{\beta
\hbar \omega _{0}}-e^{-\beta \hbar \omega _{0}})}\frac{r^{2}
}{a^{2}}\right) }\right] \right\}
\end{eqnarray}

Carrying out the inverse Laplace transformation yields
\begin{eqnarray}\label{CurrentDensityExplicit}
&&j(r)=\frac{1}{\pi a^{2}}r e^{-r^2/a^{2}}\sum\limits_{k=0}^{\infty }\sum\limits_{n=0}^{\infty
}\sum\limits_{m=0}^{\infty }\frac{1}{n!m!}\left(r^2/a^{2}\right) ^{n+m}\times  \notag \\
&&\left\{ \omega _{0}\left\{ \Theta \left[ \mu -2k\hbar \omega
_{0}-E_{n,m}-\hbar (\omega _{0}-\Omega )\right] -\Theta \left[ \mu
-2k\hbar \omega _{0}-E_{n,m}-\hbar (\omega _{0}+\Omega )\right] \right\}
L_{k}^{n+m+1}\left( 2r^2/a^{2}\right) \right.  \notag \\
&&\left. -\Omega \Theta \left[ \mu -2k\hbar \omega _{0}-E_{n,m}\right]
L_{k}^{n+m}\left( 2r^2/a^{2}\right) \right\} 
\end{eqnarray}
\end{widetext}

The explicit formulas~(\ref{DensityExplicit}) and~(\ref{CurrentDensityExplicit}) can be evaluated numerically with a high degree of accuracy.  As an example, in Fig.~\ref{Fig1} the density and current profiles obtained in this manner have been plotted in units of $1/\pi a^{2}$ and $\omega _{0}/\pi a$ respectively,  for  $\Omega /\omega _{0} =0.995$  and  $N=2000$ fermions. 


With the formation of  incompressible and compressible
regions at very fast rotation, the current flow exhibits a distinctive
pattern of reversals. As can clearly be seen in Fig.~\ref{Fig1}, the current density
flows in opposite directions in these two types of regions.
In particular, we can show~\cite{[21]} that the current in the incompressible regions in the regime of fast rotation has the form
\begin{equation}\label{JBULK}
\vec j_{bulk}\left(\vec r\right) =\frac{1}{2M^{\ast }\omega _{0}}\left(\hat z \times 
\vec \nabla V_{eff}\right) \rho (\vec r)\,,
\end{equation} 
which proportional to the gradient of the effective confinement potential.  This has the structure of a quantum Hall current.  Whereas, in the compressible regions we have
\begin{equation}\label{JEDGE}
\vec j_{edge}(\vec r) =\frac{\hbar}{2M^*}\sum_{n=0}^{\infty}\left(2n+1\right)
\left(\hat z \times\vec \nabla \rho_n(\vec r)\right)\,,
\end{equation}
with $\rho=\sum_{n=0}^\infty \rho_n$ and $\rho_n(\vec r)\simeq \frac{1}{\pi\ell^2}\Theta[\mu-V_{eff}(\vec r)-(2n+1)\hbar\omega_0]$,
which is proportional to the gradient of the particle density and therefore has the structure of an edge current.

The striking inversions in the direction of circulation of the current density are related, via equation~(\ref{DefM}),
to the oscillatory behavior of  the pseudo-magnetization $M$ as a function of $r$.
The physics of these oscillations is essentially the same as the physics of the de Haas-van Alphen effect~\cite{Alphen32} for electronic systems in a magnetic field.  In the limit in which the density is slowly-varying (on the scale of the pseudo-magnetic length $\ell$) the pseudo-magnetization is simply a function of the local density, and this function has the same form that it would have in an {\it infinite} uniform gas at the same density in the presence of the magnetic field $B=2M^*\Omega/e$.  Thus we have  $j(r)=-(\partial M/\partial \rho)_B \rho'(r)$ where the partial derivative is taken at constant $B$.  Now, according to standard thermodynamic relations, the derivative of the magnetization of a uniform gas (in the thermodynamic limit)  with respect to density at constant magnetic field  equals the derivative of the density with respect to magnetic field at constant chemical potential.  Thus we have 
\begin{equation}
j(r)\simeq -\left.\frac{\partial \rho}{\partial B}\right\vert_{\mu} \rho'(r)\,.
\end{equation}
The thermodynamic derivative $(\partial \rho/\partial B)_\mu$ is the same that appears in the Streda formula~\cite{Streda82} for the quantum Hall conductance: this quantity is positive when $\mu$ lies in a gap between Landau levels (i.e., in the incompressible regions) because the increasing $B$ increases the degeneracy of each Landau level.  But it becomes negative whenever $\mu$  crosses  a Landau level (compressible region) for, in this case, the density must decrease to compensate the upward shift of the Landau level energy with increasing $B$.  Hence we see that an oscillatory behavior of $\rho$ as a function of $B$ is responsible for the oscillations of $M$ as a function of $\rho$, hence for the reversals in the direction of flow of the current, which we observe in Fig.~\ref{Fig1}.
Although the above arguments are suggestive, we emphasize that our calculations are in no way dependent on local approximations or other heuristic arguments: they are exact results following  from the solution of the Schr\"odinger equation for noninteracting particles.
We believe that rapidly rotating  gases of Fermionic atoms offer an ideal opportunity to verify, for the first time, the striking spatial distribution of  persistent quantum currents.

GV acknowledges support from NSF Grant DMR-1104788.


\end{document}